\documentclass{revtex4}
\usepackage{amsmath} 
\usepackage{amsfonts} 

\begin{document}

\title{Relativistic mass and modern physics}
\author{ Z.~K.~Silagadze}
\affiliation{Budker Institute of Nuclear Physics SB RAS and Novosibirsk State
University, 630 090, Novosibirsk, Russia }
\email{silagadze@inp.nsk.su}

\begin{abstract}
At first sight, arguments for and against the notion of relativistic mass 
look like a notorious intra-Lilliputian quarrel between Big-Endians (those 
who broke their eggs at the larger end) and Little-Endians. However, upon
closer inspection we discover that the relativistic mass notion is alien
to the spirit of modern physics to a much greater extent than 
it seems. To demonstrate an abyss between the modern approach and archaic 
notions, in this paper we explore how the concept of mass is introduced 
in modern physics. This modern approach reveals a deep cohomological origin 
of mass.
\end{abstract}

\maketitle

\section{Introduction}

The notion of velocity-dependent ``relativistic mass'' is still used in 
the teaching of relativity and especially in the popular literature. Several 
authors have criticized the use of this concept as historically outdated 
\cite{1,2,3,4,5,6,7}, while others find it useful \cite{8,9,10,11,12,12P}. 

The concept of relativistic mass and ``Einstein's most famous equation 
$E=mc^2$'' \cite{13} were quite common in old textbooks. So why should
we challenge it? If such renowned experts in the field as Tolman, Born, 
and Fock in the past and Penrose and Rindler today find the concept of 
relativistic mass useful, why not to follow the motto ``All true believers 
break their eggs at the convenient end'' \cite{14}, instead of entering in 
an endless and arid dispute between Big-Endians and Little-Endians?

The answer is simple. Modern physics offers a picture of reality that is
completely different from the classical Newtonian picture. It is impossible
to master this kind of reality if you try to put it in a Procrustean bed
of Newtonian concepts. Nevertheless this is exactly what some modern educators
are trying to do. This effort is not unique to the realm of special 
relativity. ``Most 
elementary textbooks and popularization works about quantum physics remain 
plagued by archaic wordings and formulations'' \cite{15}.

Modern civilization depends on advances in science more than ever before.
It is of crucial importance to  cultivate conceptual critical thinking skills
in current practice of teaching and avoid authoritarian teaching traditions
\cite{16}.

The concept of mass in modern physics is quite different from the Newtonian
concept of mass as a measure of inertia. However, this does not mean that we 
should throw out mass as a measure of inertia. The modern physics framework
is more general and flexible, and it explicitly indicates the context under 
which it is fairly safe to consider the mass as a measure of inertia. The 
problems occur when this situation is turned upside-down so that Newtonian 
physics is
considered as a basic truth and modern physics as some derivative from it.
``Objectivity of Classical physics is some sort of half-truth. It is a very
good thing, a very great achievement, but somehow it makes it more difficult 
than it would have seemed before to understand the fullness of reality'' 
\cite{17}.

In this note we outline how the concept of mass is used in 
modern physics and make it clear that this use leaves no room for 
velocity-dependent ``relativistic 
mass''. That archaic concept must be discarded if we want new generations 
to fully appreciate the benefits 
of the twentieth-century scientific revolution. Our goal is to focus 
the reader's attention on a real problem. The problem is that modern 
education lags far behind the frontier of modern physics.

\section{The Landau \& Lifshitz way of introducing mass}

Okun remarks \cite{19} that the first textbook in the world in which mass was 
velocity-independent was  \textit{The Classical Theory of Fields} by 
Landau and 
Lifshitz, first published in 1940. There was a good reason why Landau and 
Lifshitz did not use relativistic mass: they had based their presentation on 
the principle of least action. And this method leaves little room for 
relativistic mass, or, rather, makes its use obsolete and unnecessary.

Let us consider a free relativistic particle. Landau and Lifshitz's
reasoning goes as follows \cite{20}. The action integral for this particle 
should be independent of our choice of reference frame, according to the 
principle of relativity. Furthermore, the integrand should be a differential 
of the first order. A free particle can provide only one scalar of this
kind: the invariant interval $ds=\sqrt{c^2\,dt^2-d{r}^{2}}$. Therefore, for 
a free particle the expected form of the action is
\begin{equation}
S=-\alpha\int\limits_a^b ds= -\alpha c \int\limits_{t_1}^{t_2}\sqrt{1-
\frac{v^2}{c^2}}\; dt
\label{eq1}
\end{equation}
where $c$ is the light velocity and the first integral is along the world line 
of the particle, $a$ and $b$ being the events defined by the arrival of the 
particle at the initial and final positions at definite times $t_1$ and $t_2$. 
The constant $\alpha$ must be positive, lest the action become 
unbounded from below.

The physical meaning of the constant $\alpha$ becomes evident if we consider
the nonrelativistic limit of the relativistic Lagrangian from Eq.~(\ref{eq1}):
\begin{equation}
L=-\alpha c \,\sqrt{1-\frac{v^2}{c^2}}\approx \alpha\, \frac{v^2}{2c}-\alpha c.
\label{eq2}
\end{equation}
This is equivalent to the nonrelativistic Lagrangian, $L=\frac{mv^2}{2}$, 
if and only if $\alpha=mc$.

The way in which the mass $m$ has appeared in the relativistic Lagrangian 
indicates clearly that it is an invariant quantity and does not depend on 
velocity. But what about $p=mv$, which is notoriously used to introduce 
relativistic mass? The simple expression $p=mv$ is a wrong way to define 
momentum. It is correct only in nonrelativistic situations and is no longer 
valid when the particle velocity approaches the velocity of light $c$.

The modern way to introduce momentum is the Noether theorem, which relates 
the symmetries of a theory to its conservation laws. In relativity, it is
better to put the time $t$ and spatial coordinates $x_i$ on equal 
footing. We therefore introduce a parametrization of the particle's 
worldline,
\begin{equation}
x_i=x_i(s),\qquad t=t(s),
\end{equation}
where $s$ is some (scalar) evolution parameter, and rewrite the action
of a dynamical system as
\begin{equation}
S=\int\limits_{t_1}^{t_2} L(x_i,v_i,t)\,dt=\int\limits_{s_1}^{s_2}
L\bigl(x_i,\frac{\dot{x}_i}{\dot{t}},t\bigr)\dot{t}\,ds=
\int\limits_{s_1}^{s_2}{\cal L}(x_i,t,\dot{x}_i,\dot{t})\,ds,
\label{eq3}
\end{equation}
\noindent where
\begin{equation}
\dot{x}_i=\frac{dx_i}{ds},\quad \dot{t}=\frac{dt}{ds},\quad
v_i=\frac{dx_i}{dt}=\frac{\dot{x}_i}{\dot{t}},
\label{eq3P}
\end{equation}
and
\begin{equation}
{\cal L}=L\bigl(x_i,\frac{\dot{x}_i}{\dot{t}},t\bigr)\dot{t}.
\label{eq3PP}
\end{equation}
A symmetry of the action (\ref{eq3}) is a transformation
\begin{equation}
x_i\to x^\prime_i=x_i+\delta x_i,\qquad t\to t^\prime=t+\delta t,
\label{eq4}
\end{equation}
under which  the variation of the Lagrangian can be written as a total 
derivative of some function $F(x_i,t,s)$ with respect to the evolution 
parameter $s$,
\begin{equation}
\delta {\cal L}=\frac{dF}{ds}.
\label{eq5}
\end{equation}
If a dynamical  system with action (\ref{eq3}) has a symmetry defined by 
(\ref{eq4}) and (\ref{eq5}), then the Noether's current (written here using
the Einstein summation convention),
\begin{equation}
J=\frac{\partial {\cal L}}{\partial \dot{x}_i}\,\delta x_i+
\frac{\partial {\cal L}}{\partial \dot{t}}\,\delta t -F,
\label{eq6}
\end{equation}
is conserved. Indeed, using the Euler-Lagrange equations that follow from 
the principle of least action $\delta S=0$, 
\begin{equation}
\frac{d}{ds}\left (\frac{\partial {\cal L}}{\partial \dot{x}_i}\right)=
\frac{\partial {\cal L}}{\partial x_i},\qquad
\frac{d}{ds}\left (\frac{\partial {\cal L}}{\partial \dot{t}}\right)=
\frac{\partial {\cal L}}{\partial t},
\end{equation}
we obtain, using Eq.~(\ref{eq5}),
\begin{equation}
\frac{dJ}{ds}=\frac{\partial {\cal L}}{\partial x_i}\delta x_i+
\frac{\partial {\cal L}}{\partial t}\delta t +
\frac{\partial {\cal L}}{\partial \dot{x}_i}\delta \dot{x}_i+
\frac{\partial {\cal L}}{\partial \dot{t}}\delta \dot{t}-\frac{dF}{ds}=
\delta {\cal L}-\delta {\cal L}=0.
\end{equation}

Sometimes it is more convenient to express the conserved Noether current in 
terms of the original Lagrangian $L$. Since
\begin{equation}
{\cal L}(x_i,t,\dot{x}_i,\dot{t})=L(x_i,v_i,t)\,\dot{t},
\end{equation}
where
\begin{equation}
v_i=\frac{\dot{x}_i}{\dot{t}},
\end{equation}
we have
\begin{equation}
\frac{\partial {\cal L}}{\partial \dot{x}_i}=\dot{t}\,\frac{\partial L}
{\partial v_j}\,\frac{\partial}{\partial \dot{x}_i}\left (\frac{\dot{x}_j}
{\dot{t}}\right )=\frac{\partial L}{\partial v_i}
\end{equation}
and
\begin{equation}
\frac{\partial {\cal L}}{\partial \dot{t}}=L+\dot{t}\,\frac{\partial L}
{\partial v_i}\,\frac{\partial}{\partial \dot{t}}\left (\frac{
\dot{x}_i}{\dot{t}}\right )=L-\frac{\partial L}{\partial v_i}\,\frac{
\dot{x}_i}{\dot{t}}=L-v_i\,\frac{\partial L}{\partial v_i}.
\end{equation}
Therefore, in terms of the original Lagrangian $L$, the Noether current 
takes the form \cite{21}
\begin{equation}
J=\frac{\partial L}{\partial v_i}\,\delta x_i-\left( v_i\,\frac{\partial L}
{\partial v_i}-L\right )\delta t -F=p_i\,\delta x_i-H\,\delta t-F,
\label{eq7}
\end{equation}
where
\begin{equation}
p_i=\frac{\partial L}{\partial v_i}\quad \mathrm{and}\quad
H=v_i\,\frac{\partial L}{\partial v_i}-L
\label{eq8}
\end{equation}
are the desired general definitions of momentum and energy (Hamiltonian)
of the dynamical system. (For the nonrelativistic Lagrangian $L=mv_iv_i/2$,
the momentum takes its standard form $p_i=mv_i$ and $H=mv_iv_i/2=p_ip_i/(2m)$
is the kinetic energy.)

It is now clear that the symmetry of the free relativistic particle action, 
Eq.~(\ref{eq1}), with respect to the infinitesimal space translations,
\begin{equation}
x^\prime_i=x_i+\epsilon_i,\quad t^\prime=t,\quad F=0,
\end{equation}
leads to momentum conservation, while the symmetry with respect to
infinitesimal time translation,
\begin{equation}
x^\prime_i=x_i,\quad t^\prime=t+\epsilon,\quad F=0,
\end{equation}
implies energy conservation.

From Eqs.~(\ref{eq8}), using the relativistic Lagrangian
\begin{equation}
L=-mc^2\sqrt{1-\frac{{v}^{2}}{c^2}},
\label{eq9}
\end{equation}
we get the relativistic expressions for the energy and momentum:
\begin{equation}
E=\frac{mc^2}{\sqrt{1-\frac{{v}^{2}}{c^2}}},\qquad
\vec{p}=\frac{m\vec{v}}{\sqrt{1-\frac{{v}^{2}}{c^2}}}.
\label{eq10}
\end{equation}
It follows from these equations that
\begin{equation}
\frac{E^2}{c^2}-{p}^{2}=m^2c^2.
\label{eq11}
\end{equation}
This equation expresses the most important relativistic facet
of mass: for every free particle, the energy-momentum four-vector has a fixed
magnitude $mc$.

An interesting question, not usually discussed in classical mechanics 
textbooks, is what conserved quantity corresponds to Lorentzian (and Galilean)
boosts \cite{22,23,24,25,26}. An infinitesimal Lorentz boost in the $x$ 
direction,
\begin{equation}
t^\prime=t-\frac{\epsilon}{c^2}\,x,\quad x^\prime=x-\epsilon t,
\quad y^\prime=y,\quad z^\prime=z,
\end{equation}
is a symmetry of the action (\ref{eq3}) with $F=0$. Therefore the 
corresponding Noether current implies the conserved quantity
\begin{equation}
p_x t-\frac{E}{c^2}x=\mathrm{constant},
\end{equation}
or in vector form, after the invariance with respect to the other two 
boosts is also taken into account,
\begin{equation}
\vec{p}\, t-\frac{E}{c^2}\,\vec{r}=\mathrm{constant}.
\label{eq12}
\end{equation}
Thus we obtain a relativistic version of the Newton's first law, that a free
particle moves uniformly with constant velocity
\begin{equation}
\vec{v}=\frac{\vec{p}c^2}{E}.
\label{eq13}
\end{equation}
``You aren't used to calling this a conservation law, but it is, and in fact
it is the Lorentz partner of the angular momentum conservation law'' 
\cite{25}.

\section{Mass, cocycles, and central extensions}

Landau and Lifshitz's argument for the form of $L$ and the invariance of $m$ 
is seductive, but only half true. Let's see how 
they get the nonrelativistic free particle  Lagrangian \cite{27}. Homogeneity 
of space and time implies that $L$ must be independent of $\vec{r}$ and $t$, 
so it is a function of the particle's velocity $\vec{v}$ and in fact a 
function of its magnitude only, because space is isotropic. Under the 
infinitesimal Galilei transformations,
\begin{equation}
x^\prime_i=x_i-\epsilon_i t,\qquad t^\prime=t,
\label{eq14}
\end{equation}
the Lagrangian $L(v^2)$ can get, at most, a variation that is a total time
derivative of some function of coordinates and time (that is, Galilean boosts
are symmetries of the corresponding action with possibly nonzero~$F$). But
under the Galilei boost (\ref{eq14}),
\begin{equation}
v^\prime_i=v_i-\epsilon_i
\end{equation}
and
\begin{equation}
\delta L=\frac{dL}{dv^2}2v_i\delta v_i=-2v_i\epsilon_i \frac{dL}{dv^2}.
\end{equation}
This is a total time derivative if and only if $dL/dv^2$ is a 
constant. Therefore we can write the Lagrangian as 
\begin{equation}
L=\frac{m}{2}\,{v}^{2}.
\label{eq15}
\end{equation}
Once again, the derivation makes it clear that the mass $m$ is 
a Galilean-invariant quantity, independent of velocity.

However, unlike the relativistic case, only quasi-invariance of the Lagrangian
is required under Galilei transformations (i.e., that the variation of the 
Lagrangian should be a total time derivative). Why such a difference? It is 
not that we cannot find a Lagrangian that is invariant under Galilean 
transformations (\ref{eq14}). We can. Just adding a total time derivative
to the Lagrangian (\ref{eq15}), we get a Lagrangian $\tilde L$ that is 
evidently invariant under Galilean boosts \cite{26}:
\begin{equation}
\tilde L=L+\frac{d}{dt}\left (-\frac{m{r}^{2}}{2t}\right )=
\frac{m}{2}\left |\vec{v}-\frac{\vec{r}}{t}\right |^2.
\label{eq16}
\end{equation}
Of course, $\tilde L$ explicitly depends on $\vec{r}$ and $t$. But this
fact, contrary to what is claimed in Ref.~\cite{27}, does not mean 
a violation of space-time homogeneity. Under space-time translations,
\begin{equation}
t^\prime=t+\tau,\qquad \vec{r}^{\,\prime}=\vec{r}+\vec{a},
\end{equation}
the variation of $\tilde L$ is
\begin{equation}
\delta \tilde L=\frac{d}{dt}\left [ \frac{m}{2}\left ( \frac{{r}^{2}}
{t}- \frac{|\vec{r}+\vec{a}|^2}{t+\tau}\right )\right ].
\end{equation}
Therefore, the Lagrangian $\tilde L$ is quasi-invariant under space-time 
translations and this is sufficient to ensure space-time homogeneity.

As we see, Landau and Lifshitz's logic, while yielding free particle 
Lagrangians in Refs.~\cite{20} and~\cite{27}, contains loopholes. 
To close these loopholes, more thorough investigation is necessary \cite{28} 
(see also Refs.~\cite{26}, \cite{29}, and~\cite{29P}).

For notational simplicity, let $q$ denote a space-time point $(t(s),\vec{r}
(s))$, and let the Lagrangian ${\cal L}(q,\dot q)$ be quasi-invariant with
respect to the symmetry group $G$. That is, for any symmetry transformation
$g\in G$, we have
\begin{equation}
{\cal L}(gq,g\dot q)={\cal L}(q,\dot q)+\frac{d}{ds}\alpha(g;q).
\label{eq17}
\end{equation}
The action
\begin{equation}
S(q_1,q_2)=\int\limits_{s_1}^{s_2} {\cal L}(q,\dot q)\,ds,
\label{eq18}
\end{equation}
considered as a function of the trajectory end-points, transforms as
\begin{equation}
S(gq_1,gq_2)=S(q_1,q_2)+\alpha(g;q_2)-\alpha(g;q_1).
\label{eq19}
\end{equation}
L\'{e}vy-Leblond calls $\alpha(g;q)$ a gauge function. If this function has
the form 
\begin{equation}
\alpha(g;q)=\phi(q)-\phi(gq)+\chi(g),
\label{eq20}
\end{equation}
with some functions $\phi$ and $\chi$, then we can choose a new equivalent 
action 
\begin{equation}
\tilde S(q_1,q_2)=S(q_1,q_2)+\phi(q_2)-\phi(q_1),
\end{equation}
which will be invariant under all symmetry transformations from $G$ (this
follows simply from Eqs.\ (\ref{eq19}) and~(\ref{eq20})):
\begin{equation}
\tilde S(gq_1,gq_2)=\tilde S (q_1,q_2).
\end{equation}
In this case the gauge function $\alpha(g;q)$ is said to be equivalent to 
zero. Of course, two gauge functions are essentially the same (are equivalent) 
if their difference is equivalent to zero. It is, therefore, convenient to fix 
the gauge and choose one representative from each equivalence class with the
property
\begin{equation}
\alpha(g;q_0)=0\quad \mathrm{for\ any}\ g\in G, 
\label{eq21}
\end{equation}
where $q_0$ denotes a conventional origin $\vec{r}=0,\,t=0$ in space-time
(of course, any point can be chosen as the origin, due to space-time 
homogeneity). Such a representative always exists because if $\alpha(g;q_0)
\ne 0$, we choose as a representative an equivalent gauge function
$\tilde \alpha(g;q)=\alpha(g;q)-\chi(g)$, with $\chi(g)=\alpha(g;q_0)$.

The gauge functions have the following important property \cite{28}.
The compatibility of
\begin{equation}
S(g_1g_2q_1,\,g_1g_2q_2)= S(q_1,q_2)+\alpha(g_1g_2;q_2)-
\alpha(g_1g_2;q_1)
\end{equation}
and
\begin{equation}
S(g_1g_2q_1,\,g_1g_2q_2) = 
S(g_2q_1,\,g_2q_2)+\alpha(g_1;g_2q_2)
-\alpha(g_1;g_2q_1)=
S(q_1,q_2)+\alpha(g_2;q_2)- 
\alpha(g_2;q_1)+
\alpha(g_1;g_2q_2)-\alpha(g_1;g_2q_1) 
\end{equation}
requires
\begin{equation} 
\alpha(g_2;q_1)+\alpha(g_1;g_2q_1)-\alpha(g_1g_2;q_1)= 
\alpha(g_2;q_2)+\alpha(g_1;g_2q_2)-\alpha(g_1g_2;q_2).
\end{equation}
That is, the function 
\begin{equation}
\xi(g_1,g_2)=\alpha(g_2;q)+\alpha(g_1;g_2q)-\alpha(g_1g_2;q) 
\label{eq22}
\end{equation}
does not depends on the space-time point $q$.

Some elementary cohomology terminology will be useful at this point \cite{30}. 
Cohomological methods are powerful but sophisticated tools applicable 
ubiquitously in modern mathematics. Unfortunately what follows may seem
too abstract for phy\-sics oriented readers. In this case we recommend to 
consult Refs.~\cite{30A} and \cite{30B} in which Kirchoff's 
results on electric circuits are presented in a way to motivate an 
introduction of cohomological notions. Other great source of inspiration is
Ref.~\cite{30C} which demonstrates that carrying in manual addition
of two multi-digit numbers is a particular example of cocycle condition
and uses this fact to illustrate some aspects of group cohomology.

Any real function $\alpha_n(g_1,g_2,\ldots,g_n;q)$ will be called an 
$n$-cochain. 
The action of the coboundary operator $\delta$ on this $n$-cochain produces
an $(n+1)$-cochain, defined as follows:
\begin{eqnarray} 
&&(\delta \alpha_n)(g_1,g_2,\ldots,g_n,g_{n+1};q) = 
\alpha_n(g_2,g_3,\ldots,g_n,g_{n+1};g_1^{-1}q)-\alpha_n(g_1\cdot g_2,
g_3,\ldots,g_n,g_{n+1};q)+ \nonumber \\ 
&& \qquad\alpha_n(g_1,g_2\cdot g_3,g_4,\ldots,g_n,g_{n+1};q)-
\alpha_n(g_1,g_2,g_3\cdot g_4,g_5,\ldots,g_n,g_{n+1};q)+\nonumber \\ 
&& \qquad\cdots+(-1)^{n}\alpha_n(g_1,g_2,\ldots,g_n\cdot g_{n+1};q)+(-1)^{n+1}
\alpha_n(g_1,g_2,\ldots,g_n;q). \label{eq23}
\end{eqnarray}
\noindent The coboundary operator has the important property
\begin{equation}
\delta^2=0.
\label{eq24}
\end{equation}
A cochain with zero coboundary is called a cocycle. Because of 
Eq.~(\ref{eq24}),
every coboundary $\alpha_n=\delta \alpha_{n-1}$ is a cocycle. However, not 
all cocycles can be represented as coboundaries; such cocycles will be called 
nontrivial.

In fact, $\xi(g_1,g_2)$ defined by Eq.~(\ref{eq22}) is a cocycle. For according
to Eq.~(\ref{eq23}),
\begin{equation} 
(\delta \xi)(g_1,g_2,g_3;q)= 
\xi(g_2,g_3)-\xi(g_1g_2,g_3)+\xi(g_1,g_2g_3)-
\xi(g_1,g_2).
\end{equation}
Substituting Eq.~(\ref{eq22}) into the first three terms, we get after some 
cancellations
\begin{equation} 
\xi(g_2,g_3)-\xi(g_1g_2,g_3)+\xi(g_1,g_2g_3)= 
\alpha(g_2;g_3q)+
\alpha(g_1;g_2g_3q)-\alpha(g_1g_2;g_3q),
\end{equation}
but this is just $\xi(g_1,g_2)$, as the formula (\ref{eq22}) is valid for any
space-time point $q$, and in particular for the point  $p=g_3q$. Therefore, 
$(\delta \xi)(g_1,g_2,g_3;q)=0$ and $\xi(g_1,g_2)$ is a global (independent of 
any space-time point $q$) cocycle.

In the following we will assume the gauge fixing condition (\ref{eq21}). Then  
Eq.~(\ref{eq22}) with $q=q_0$ gives
\begin{equation}
\xi(g_1,g_2)=\alpha(g_1;g_2q_0).
\label{eq25}
\end{equation}
From this relation the following two properties of the admissible cocycles
follow. 

First, if the gauge function $\alpha(g;q)$ is equivalent to zero,
then $\xi(g_1,g_2)$ is a trivial cocycle. Indeed, let
\begin{equation}
\alpha(g;q)=\phi(q)-\phi(gq)+\chi(g).
\end{equation}
Then the condition $\alpha(g;q_0)=0$ gives 
\begin{equation}
\chi(g)=\phi(gq_0)-\phi(q_0),
\end{equation}
and, therefore,
\begin{equation}
\xi(g_1,g_2)=\phi(g_2q_0)-\phi(g_1g_2q_0)+\phi(g_1q_0)-\phi(q_0).
\label{eq26}
\end{equation}
On the other hand, if we take (a global) 1-cochain $\beta(g)=\phi(gq_0)-
\phi(q_0)$, then its coboundary
\begin{equation}
(\delta \beta)(g_1,g_2)=\beta(g_2)-\beta(g_1g_2)+\beta(g_1)
\end{equation}
just coincides with the right-hand side of Eq.~(\ref{eq26}). Therefore, 
$\xi=\delta \beta$ and hence it is a trivial cocycle.

The second property of the admissible cocycle is that if $h\in \Gamma$ 
belongs to the stabilizer $\Gamma$ of the point $q_0$, so that $hq_0=q_0$,
then $\xi(g,h)=0$ for all $g\in G$. This is evident from Eq.~(\ref{eq25})
and our gauge fixing condition (\ref{eq21}).

If the symmetry group $G$ acts transitively on space-time (in fact, in this 
case the space-time can be identified with the homogeneous space $G/\Gamma$; 
see Ref.~\cite{28}), then for any point $q$ there exists a symmetry 
$g_q$ that
\begin{equation}
q=g_qq_0.
\label{eq27}
\end{equation}  
Let $\xi(g_1,g_2)$ be some admissible cocycle, such that $\xi(g,h)=
0$ for all $g\in G$ and $h\in \Gamma$. Then the formula
\begin{equation}
\alpha(g;q)=\xi(g,g_q)
\label{eq28}
\end{equation} 
defines a gauge function such that $\alpha(g;q_0)=0$. Indeed, first of all
$\alpha(g;q)$ is defined by Eq.~(\ref{eq28}) for admissible cocycles uniquely, 
despite the fact that $g_q$ is defined by Eq.~(\ref{eq27}) only up to 
stabilizer transformation. Namely, for any $h\in \Gamma$ we have 
\begin{equation} 
\xi(g_1,g_2h)=(\delta \xi)(g_1,g_2,h)+\xi(g_1,g_2)+ 
\xi(g_1g_2,h)-\xi(g_2,h)=\xi(g_1,g_2).
\end{equation}
Then, using $g_{gq}=gg_qh$ for some $h\in \Gamma$, we can easily check
that 
\begin{equation} 
\alpha(g_2;q)+\alpha(g_1;g_2q)-\alpha(g_1g_2;q)= 
(\delta \xi)(g_1,g_2,g_q)+\xi(g_1,g_2)=\xi(g_1,g_2).
\end{equation}

The only question that remains is whether the equivalent admissible cocycles
can lead to nonequivalent gauge functions. The answer, in general, turns out
to be affirmative \cite{28}.

Let $\xi^\prime(g_1,g_2)$ and $\xi(g_1,g_2)$ be two equivalent admissible 
cocycles, so that
\begin{equation}
\xi^\prime(g_1,g_2)=\xi(g_1,g_2)+\zeta(g_2)-\zeta(g_1g_2)+\zeta(g_1).
\end{equation}
The admissibility condition $\xi^\prime(g,h)=\xi(g,h)=0$, if $h\in \Gamma$,
produces a restriction  on the cochain $\zeta(g)$:
\begin{equation}
\zeta(gh)=\zeta(g)+\zeta(h),\;\;\;\mathrm{for\;\;any}\;\;g\in G\;\;
\mathrm{and}\;\;h\in\Gamma.
\label{eq29}
\end{equation}
In particular, Eq.~(\ref{eq29}) shows that $h\to\zeta(h)$ is a 
one-dimensio\-nal representation of the subgroup $\Gamma$.

The gauge functions $\alpha^\prime(g;q)$ and $\alpha(g;q)$ defined by these
cocycles are related as follows:
\begin{equation}
\alpha^\prime(g;q)=\alpha(g;q)+\zeta(g)+\zeta(g_q)-\zeta(gg_q).
\end{equation}
Note that $gg_q=g_{gq}h$ with some $h\in\Gamma$ (in fact, 
$h=g^{-1}_{gq}gg_q$). Therefore, in light of Eq.~(\ref{eq29}), we have
\begin{equation}
\alpha^\prime(g;q)=\alpha(g;q)+\zeta(g)+\zeta(g_q)-\zeta(g_{gq})-\zeta(h),
\end{equation}
or
\begin{equation}
\alpha^\prime(g;q)=\alpha(g;q)-\zeta(g^{-1}_{gq}gg_q)+\phi(q)-\phi(gq)+
\chi(g),
\end{equation}
with $\phi(q)=\zeta(g_q)$ and $\chi(g)=\zeta(g)$. As we see, 
$\alpha^\prime(g;q)$ is equivalent to the gauge function
\begin{equation}
\tilde\alpha(g;q)=\alpha(g;q)-\zeta(g^{-1}_{gq}gg_q)=
\xi(g,g_q)-\zeta(g^{-1}_{gq}gg_q).
\label{eq30}
\end{equation}
Suppose the representation $\zeta$ of $\Gamma$ can be extended to the 
representation $\omega$ of the whole group $G$. Then we will have
\begin{equation}
\zeta(g^{-1}_{gq}gg_q)=\omega(g^{-1}_{gq}gg_q)=\omega(g)+\omega(g_q)-
\omega(g_{gq}),
\end{equation}
and
\begin{equation}
\alpha(g;q)=\tilde\alpha(g;q)+\phi(q)-\phi(gq)+\chi(g),
\end{equation}
with $\phi(q)=\omega(g_q)$ and $\chi(g)=\omega(g)$. Therefore, $\alpha$ and
$\tilde\alpha$ are equivalent.

However, if the representation $\zeta$ cannot be extended on $G$, then
the gauge functions $\tilde\alpha(g;q)$ and $\alpha(g;q)$ are essentially 
different (not equivalent).

In fact, formula (\ref{eq30}) makes it possible to explicitly construct all
different gauge functions related to the symmetry group $G$. All that is 
needed is to find all nontrivial 2-cocycles of $G$ and all nontrivial 
one-dimensional representations of the stabilizer subgroup $\Gamma$ that
cannot be extended on $G$ \cite{28}.

In relativistic classical mechanics, the symmetry group $G$ is the 
Poincar\'{e} group and the stabilizer subgroup $\Gamma$ is the (homogeneous)
Lorentz group. However, the Poincar\'{e} group has no nontrivial 
2-cocycles \cite{31,32} and the Lorentz group has no nontrivial 
one-dimensional representations. Therefore, all gauge functions related
to the Poincar\'{e} group are equivalent to zero and we conclude that it
was quite safe for Landau and Lifshitz to assume a strictly invariant 
relativistic action integral.

In the nonrelativistic case, matters are somewhat more complicated. Now $G$ is
the Galilei group with elements 
\begin{equation}
g=(\tau,\vec{a},\vec{v},R),
\end{equation}
and it acts on the space-time points $q=(t,\vec{r})$ as follows:
\begin{equation}
gq=(t+\tau,\,R\vec{r}-\vec{v}t+\vec{a}),
\label{eq31}
\end{equation}
where $R$ symbolically denotes the rotation matrix. The stabilizer subgroup 
$\Gamma$ is the homogeneous Galilei group with elements $g=(0,0,\vec{v},R)$.
As in the relativistic case, $\Gamma$ has no nontrivial one-dimensional
representations. However, the full Galilei group $G$ has a nontrivial
2-cocycle discovered by Barg\-mann \cite{31}. Bargmann's cocycle may be 
chosen in the form \cite{28}
\begin{equation}
\xi(g_1,g_2)=m\left (\frac{1}{2}\vec{v}_1^{\,2}\,\tau_2-\vec{v}_1\cdot R_1
\vec{a_2}\right ),
\label{eq32}
\end{equation}
where $m$ is an arbitrary real number.

Possible gauge functions for the Galilei group are uniquely specified by the
equivalence classes of the Bargmann cocycle, that is, by the number~$m$. We can
take $g_q=(t,\vec{r},0,1)$, because $g_qq_0=q$ with $q_0=(0,0)$ and $q=(t,
\vec{r})$. Therefore, in accordance with Eq.~(\ref{eq30}), we obtain 
\begin{equation}
\alpha(g;q)=\xi(g,g_q)=m\left (\frac{1}{2}v^2t-\vec{v}\cdot R\vec{r}\right ).
\label{eq33}
\end{equation}
We then obtain the most general transformation law of the Lagrangian
under Galilei symmetries:
\begin{equation}
{\cal L}(gq,g\dot q)={\cal L}(q,\dot q)+m\left (\frac{1}{2}v^2\dot t-
\vec{v}\cdot R\dot{\vec{r}}\right ),
\label{eq34}
\end{equation}
where $gq$ is given by Eq.~(\ref{eq31}) and
\begin{equation}
g\dot q=(\dot t, \,R\dot{\vec{r}}-\vec{v}\dot t).
\label{eq35}
\end{equation}
Choosing $g=g_q^{-1}=(-t,-\vec{r},0,1)$, we get
\begin{equation}
{\cal L}(q_0,\dot q)={\cal L}(q,\dot q).
\end{equation}
Therefore, ${\cal L}$ does not depend on $\vec{r}$ and $t$, as was assumed by
Landau and Lifshitz. But now we have a rigorous justification of why we can 
make such a choice without loss of generality, in spite of quasi-invariance 
of the Lagrangian.

Thus ${\cal L}(q,\dot q)={\cal L}(\dot t,\dot{\vec{r}})$, and we can rewrite 
Eq.~(\ref{eq34}) as
\begin{equation}
{\cal L}(\dot t,R\dot{\vec{r}}-\vec{v}\dot t)={\cal L}(\dot t,\dot{\vec{r}}
\,)+m\left (\frac{1}{2}v^2\dot t-\vec{v}\cdot R\dot{\vec{r}}\right ).
\label{eq36}
\end{equation}
For $\vec{v}=R\dot{\vec{r}}/\dot t$, we get (note that $|R\dot{\vec{r}}|^2=
|\dot{\vec{r}}|^2$)
\begin{equation}
{\cal L}(\dot t,0)={\cal L}(\dot t,\dot{\vec{r}}\,)-\frac{m}{2}\,
\frac{|\dot{\vec{r}}|^{2}}{\dot t}.
\label{eq37}
\end{equation}
The Lagrangian ${\cal L}$ is a homogeneous function of first degree in the 
$\dot t$ and $\dot{\vec{r}}$ derivatives (see Eq.~(\ref{eq3PP})). Therefore,
${\cal L}(\dot t,0)=E_0\dot t$, where $E_0$ is some arbitrary constant, and 
from Eq.~(\ref{eq37}) we get the most general form (up to equivalence) of the 
Lagrangian compatible to the Galilei symmetry,
\begin{equation}
{\cal L}(\dot t,\dot{\vec{r}})=E_0\dot t+\frac{m}{2}\,
\frac{\dot{\vec{r}}^{\,2}}{\dot t}.
\label{eq38}
\end{equation}
Taking $s=t$, so that $\dot t=1$ and $\dot{\vec{r}}$ is the particle velocity,
we recover the standard result
\begin{equation}
{\cal L}(\dot t,\dot{\vec{r}})=E_0+\frac{m}{2}\left (\frac{d\vec{r}}{dt}
\right)^2.
\label{eq39}
\end{equation}
The rest energy $E_0$ that appears in Eq.~(\ref{eq39}) has no real significance
in classical mechanics and can be omitted from Eq.~(\ref{eq39}) without 
changing the equations of motion. Note, however, that the way that this rest 
energy was 
introduced in the theory indicates that $E_0$ has no relation with the mass
$m$ of the particle. In nonrelativistic physics (or more precisely, in 
Galilei-invariant theory), $E_0$ and $m$ are two unrelated constants 
characterizing the particle ($E_0$ being insignificant as far as 
classical mechanics is concerned).
Only in relativity are $E_0$ and $m$ related by Einstein's famous 
formula $E_0=mc^2$.  (In fact, it was Max Laue who 
produced the first general correct proof of this relation  in 1911 for 
arbitrary closed static systems, generalized by Felix Klein in 1918 to 
arbitrary closed time-dependent systems \cite{33}).

Interestingly, we can obtain a strictly invariant Lagrangian if we enlarge the
confi\-guration space of the system by introducing just one additional real
variable $\theta$. Consider the Lagrangian
\begin{equation}
\tilde {\cal L}= {\cal L}-\dot \theta.
\label{eq40}
\end{equation}
It is invariant under the transformation
\begin{equation}
q^\prime=gq,\qquad \theta^\prime=\theta+\alpha(g;q),
\label{eq41}
\end{equation}
because 
\begin{equation}
{\cal L}(q^\prime,\dot{q}^\prime)={\cal L}(q,\dot q)+\frac{d}{ds}
\alpha(g;q).
\end{equation}
Unfortunately, the transformations of the form (\ref{eq41}) do not form 
a group:
\begin{equation} 
g_1[g_2(q,\theta)] 
= (g_1g_2q,\,\theta+\alpha(g_2;q)+\alpha(g_1;g_2q)) 
= (g_1g_2q,\,\theta+\alpha(g_1g_2;q)+\xi(g_1,g_2)) 
\ne  (g_1g_2)(q,\theta)=(g_1g_2q,\,\theta+\alpha(g_1g_2;q)).
\end{equation}
As we see, the presence of the $\xi(g_1,g_2)$ cocycle makes it impossible
to define the multiplication law $g_1\odot g_2$ because 
$$g_1[g_2(q,\theta)]\ne (g_1g_2)(q,\theta).$$

However, there is a simple way out. The Lagrangian (\ref{eq40}) is
invariant under transformation
\begin{equation}
q^\prime=q,\qquad \theta^\prime=\theta+\Theta,
\label{eq42}
\end{equation}
with some constant $\Theta$. Let us combine the transformations (\ref{eq41})
and (\ref{eq42}) in the following way:
\begin{equation}
(g,\Theta)(q,\theta)=(gq,\,\theta+\Theta+\alpha(g;q)).
\label{eq43}
\end{equation}
Then the condition $g_1[g_2(q,\theta)]=(g_1g_2)(q,\theta)$ requires the
multiplication law
\begin{equation}
(g_1,\Theta_1)\odot (g_2,\Theta_2)=(g_1g_2,\,\Theta_1+\Theta_2+\xi(g_1,g_2)).
\label{eq44}
\end{equation}
It can be checked that in this case the cocycle condition
\begin{equation}
\xi(g_2,g_3)+\xi(g_1,g_2g_3)=\xi(g_1,g_2)+\xi(g_1g_2,g_3)
\end{equation}
helps to ensure the associativity of the multiplication law (\ref{eq44}) and
the set of the $(g,\Theta)$ pairs, $\tilde G$, indeed form a group, the 
inverse element being
\begin{equation}
(g,\Theta)^{-1}=(g^{-1},\,-\Theta-\xi(g,g^{-1})).
\end{equation}
Note that $G$ is not a subgroup of $\tilde G$. Instead, $G$ is isomorphic
to the factor-group $\tilde G/\mathbb{R}$, where $\mathbb{R}$ is the Abelian 
group of transformations (\ref{eq42}) (identical to the additive group of
real numbers). It is said that $\tilde G$ constitutes a central extension
of $G$. Central extensions play an important role in physics, especially in
quantum physics \cite{26,34}.

\section{Mass and quantum theory}

Although the classical theory, considered above, is comple\-tely sufficient to
demonstrate our main point that the modern concept of mass cannot depend on 
velocity in either Galilei- or Poincar\'{e}-invariant theory, the real basis 
of modern physics is quantum theory.

Through the Feynman path integral formalism, the quantum theory explains
the appearance of the least action principle in classical theory \cite{35}.
Therefore, ``there is no longer any need for the mystery that comes from 
trying to describe quantum behaviour as some strange approximation to the 
classical behaviour of waves and particles. Instead we turn the job of 
explaining around. We start from quantum behaviour and show how this explains
classical behaviour'' \cite{36}.

Unfortunately, modern education, as a rule, completely ignores this approach 
and still
speaks about ``wave-particle duality'' and ``the complementarity principle'' 
as philosophical bases to fuse the two apparently contradictory ideas of 
classical particles and classical waves into a quantum concept. The following
example \cite{37} shows that such educational practice distorts the scientific
integrity even of professional physicists. In his critique of the customary 
interpretation of quantum mechanics, Land\'{e} described the following 
paradox \cite{38}. It seems the de~Broglie relationship between the momentum 
and wavelength,
\begin{equation}
p=\frac{h}{\lambda},
\end{equation}
contradicts the principle of relativity because the momentum depends on 
the choice of the reference frame while the wavelength does not. An amusing 
fact, according to Lev\'{y}-Leblond \cite{37}, is that nowadays 
experimentalists in neutron optics have difficulty grasping Land\'{e}'s 
paradox. 
They fail to appreciate the crux of the  Land\'{e}'s argument, 
that ``in classical wave theory, $\lambda$ indeed is an invariant: the 
crest-to-crest distance of sea waves is the same to an aircraft pilot and to 
a lighthouse keeper'' \cite{37}. 

Although  Land\'{e}'s paradox has some interesting aspects related 
to classical special relativity \cite{39}, it is not a real paradox in quantum
theory \cite{40} and arises only if we still insist on the schizophrenic 
classical view of the quantum world that a quantum particle somehow manages 
to be simultaneously both a particle and a wave while in reality it is neither 
particle nor wave \cite{41}. In other words, ``it must be realized today that 
this view of the quantum world, adapted as it was to its first explorations, 
is totally out-dated. In the past fifty years, we have accumulated sufficient 
familiarity, theoretical as well as experimental, with the quantum world
to no longer look at it through classical glasses'' \cite{37}.

Mathematically, Land\'{e}'s paradox is the following. The wave function of 
a free nonrelativistic particle (for simplicity, we will assume $\hbar=1$),
\begin{equation}
\Psi(\vec{r},t)=\exp{\{i(Et-\vec{p}\cdot\vec{r})\}},
\end{equation}
is not invariant under Galilei boosts,
\begin{eqnarray} &&
t^\prime=t,\quad \vec{r}^{\,\prime}=\vec{r}-\vec{v}\,t,\nonumber \\ &&
\vec{p}^{\,\prime}=
\vec{p}-m\,\vec{v},\quad E^\prime=E-\vec{p}\cdot\vec{v}+\frac{mv^2}{2}.
\end{eqnarray}
That is, $\Psi^\prime(\vec{r}^{\,\prime},t^\prime)\ne\Psi(\vec{r},t)$, where
$$\Psi^\prime(\vec{r},t)=\exp{\{i(E^\prime t-\vec{p}^{\,\prime}\cdot
\vec{r})\}}.$$
This is of course true, but physically, strict invariance 
is not required. What is really required is invariance up to a phase factor,
\begin{equation}
\Psi^\prime(\vec{r}^{\,\prime},t^\prime)=e^{i\tilde\alpha(g;\,\vec{r},t)}\,
\Psi(\vec{r},t).
\label{eq45}
\end{equation}
The phase $\alpha(g;q)=\tilde\alpha(g;g^{-1}q)$ is not completely arbitrary; 
let us look at conditions it must satisfy. 

Consider a sequence of Galilei boosts,
\begin{equation}
q\to g_2q\to g_1(g_2q).
\label{eq46}
\end{equation}
If we write Eq.~(\ref{eq45}) as
\begin{equation}
\Psi^\prime(q)= e^{i\alpha(g;q)}\,\Psi(g^{-1}q),
\label{eq47}
\end{equation}
then we obtain for the sequence (\ref{eq46}):
\begin{equation} 
\Psi^{\prime\prime}(q)=e^{i\alpha(g_1;q)}\,\Psi^\prime(g_1^{-1}q)= 
e^{i[\alpha(g_1;q)+\alpha(g_2;g_1^{-1}q)]}\Psi(g_2^{-1}g_1^{-1}q).
\label{eq48}
\end{equation}

The transformation (\ref{eq47}) realizes
invariance with respect to Galilei boosts if wave function (\ref{eq48})
is physically indistinguishable from the wave function
\begin{equation}
\tilde\Psi^{\prime\prime}(q)=e^{i\alpha(g_1g_2;q)}\,\Psi((g_1g_2)^{-1}q),
\label{eq49}
\end{equation}
associated with the direct $q\to (g_1g_2)q$ transition. Physical 
indistinguishability means that transition amplitudes are the same,
\begin{equation}
\Psi^{\prime\prime}(q_2)[\Psi^{\prime\prime}(q_1)]^*=
\tilde\Psi^{\prime\prime}(q_2)[\tilde\Psi^{\prime\prime}(q_1)]^*,
\label{eq50}
\end{equation}
for any two space-time points $q_1$ and $q_2$. Substituting Eqs.\ (\ref{eq48}) 
and (\ref{eq49}) into Eq.~(\ref{eq50}), we find that the combination
\begin{equation}
\xi(g_1,g_2)=\alpha(g_2;g_1^{-1}q)-\alpha(g_1g_2;q)+\alpha(g_1;q)
\label{eq51}
\end{equation}
must be independent on the space-time point $q$. Note that
\begin{equation}
\xi(g_1,g_2)=(\delta\alpha)(g_1,g_2;q).
\end{equation}
Therefore $\xi(g_1,g_2)$ is a locally trivial cocycle, but globally
it is not necessarily trivial, that is, representable as the coboundary of a
global cochain (this is just the case for the Galilei group, as we shall see
soon).

In fact the Land\'{e} paradox  will be resolved if we show that de Broglie 
plane waves really induce a global cocycle (\ref{eq51}). Let us write 
Eq.~(\ref{eq47}) for de Broglie plane waves:
\begin{equation} 
\exp{\{-i(E^\prime t-\vec{p}^{\,\prime}\cdot\vec{r})\}}= 
e^{i\alpha(g;q)}\,\exp{\{-i[E t-\vec{p}\cdot(\vec{r}+\vec{v}\,t)]\}}.
\end{equation}
Then we get
\begin{equation}
\alpha(g;q)=-\frac{mv^2}{2}\,t-m\vec{v}\cdot\vec{r},
\label{eq52}
\end{equation}
and we can check that
\begin{equation}
\xi(g_1,g_2)=0
\end{equation}
for any two pure Galilei boosts $g_1$ and $g_2$. As we see, the phase factor
(\ref{eq52}) in Eq.~(\ref{eq45}) resolves the  Land\'{e} paradox.

In fact, $\xi(g_1,g_2)$ is the Bargmann cocycle, Eq.~(\ref{eq32}). \cite{42} 
This can be shown as follows. Repeating the above reasoning for the general
transformations from the Galilei group,
\begin{eqnarray} 
gq &=& (t+\tau,\,R\vec{r}-\vec{v}\,t+\vec{a}),\;\; 
g^{-1}q = (t-\tau,\,R^{-1}(\vec{r}+\vec{v}\,t-\vec{a}-\vec{v}\,\tau)), 
\nonumber \\ 
(g_1g_2)q &=& (t+\tau_1+\tau_2,\, R_1R_2\vec{r}-(\vec{v}_1+R_1\vec{v}_2)t+
\vec{a}_1+R_1\vec{a}_2-\vec{v}_1\tau_2), \label{eq53}
\end{eqnarray}
we get 
\begin{equation}
\alpha(g;q)=-\frac{mv^2}{2}\,(t-\tau)-m\vec{v}\cdot (\vec{r}-\vec{a})-
E^\prime\tau+\vec{p}^{\,\prime}\cdot \vec{a}.
\label{eq54}
\end{equation}
Only the first two terms are relevant because the last two terms give 
a function of group parameters only (independent on $\vec{r}$ and $t$)
and therefore lead to a globally trivial cocycle when substituted into
Eq.~(\ref{eq51}). Keeping only the first two terms in Eq.~(\ref{eq54}) and 
taking into account Eqs.~(\ref{eq53}), we get after some algebra
\begin{equation} 
\alpha(g_2;g_1^{-1}q)-\alpha(g_1g_2;q)+\alpha(g_1;q)= 
\frac{m}{2}v_1^2\,
\tau_2-m\vec{v}_1\cdot R_1\vec{a}_2,
\end{equation}
which is just the Bargmann cocycle (\ref{eq32}).

The way we have obtained it shows that the Bargmann cocycle is locally 
trivial (is the coboundary of the local cochain $\alpha(g;q)$). However,
it is globally nontrivial. Indeed, any globally trivial cocycle, having the 
form $\beta(g_2)-\beta(g_1g_2)+\beta(g_1)$, is symmetric in $g_1$ and $g_2$ 
on Abelian subgroups. It follows from Eqs.~(\ref{eq53}) that elements of the 
form $(0,\vec{a},\vec{v},1)$ (space translations and Galilean boosts) form an
Abelian subgroup:
\begin{equation}
(0,\vec{a}_1,\vec{v}_1,1)\cdot (0,\vec{a}_2,\vec{v}_2,1)=
(0,\vec{a}_1+\vec{a}_2,\vec{v}_1+\vec{v}_2,1).
\end{equation}
However, the Bargmann cocycle remains asymmetric on this Abelian subgroup,
$\xi(g_1,g_2)=-m\vec{v}_1\cdot\vec{a}_2$, and, therefore, it cannot be 
a trivial cocycle. Moreover, the same argument indicates that different
values of mass define inequivalent Barg\-mann cocycles because their 
difference, being asymmetric on the Abelian subgroup of space translations 
and Galilean boosts, is not a trivial cocycle. As we see, in 
nonrelativistic physics, the mass of the particle has a cohomological 
origin: it parametrizes the central extensions of the Galilei group.

In nonrelativistic physics, the mass is a primary concept and it is 
impossible to explain why only some central extensions of the Galilei group
are realized as elementary particles. Relativity brings a big change in the
conceptual status of mass. Einstein's $E_0=mc^2$ ``suggests the possibility 
of explaining mass in terms of energy'' \cite{43}. In fact, quantum 
chromodynamics already explains the origin of mass of most constituents 
of ordinary matter \cite{44}. However, ``our understanding of the 
origin of mass is by no means complete. We have achieved a beautiful and 
profound understanding of the origin of most of the mass of ordinary matter, 
but not of all of it. The value of the electron mass, in particular, remains 
deeply mysterious even in our most advanced speculations about unification 
and string theory. And ordinary matter, we have recently learned, supplies 
only a small fraction of mass in the Universe as a whole. More beautiful and 
profound revelations surely await discovery. We continue to search for 
concepts and theories that will allow us to understand the origin of mass in 
all its forms, by unveiling more of Nature's hidden symmetries'' \cite{43}.

\section{Concluding remarks}

V.~A.~Fock once remarked that ``physics is essentially a simple science. 
The main problem in it is to understand which symbol means what'' \cite{45}.
As we have seen above, the meaning of the symbol $m$ in Newton's
$\vec{F}=m\vec{a}$ is more profound than the primary Newtonian
``measure of inertia.'' Unfortunately, modern education ignores the 
twentieth  century's achievements in deciphering this symbol and bases 
its exposition on classical Newtonian physics as it was understood at the end 
of the nineteenth century, with only fragmentary and eclectic inclusions
from modern physics.


But all the sparkling beauty of classical physics can manifest itself only 
when it is placed in a right framework of modern ideas \cite{46}. 
Archaic notions and 
concepts in education, like relativistic mass, not only hinder understanding
of modern physics but also make it impossible to truly appreciate the 
meaning of classical ideas and the context under which classical ideas are
completely sound and operational.

Of course we are not talking about mere terminology. If you like to have
a special name for the combination $m\gamma$, or you feel that this concept 
will help your students to better understand some relativistic circumstances 
using their Newtonian intuition, then there is no reason not to go ahead and 
use it. After all, relativistic mass, if properly used, can even offer 
interesting insights in hyperbolic geometry \cite{11}. It is the philosophy 
of teaching that is at stake. 

Modern education can no longer be based on Newton's laws and Newtonian
concepts as primary building blocks. The prog\-ress in science has been too 
great. Quantum mechanics and special relativity are cornerstones of modern 
physics. It is of crucial importance that modern education be based on the 
basic principles of these disciplines from the very beginning. Newton's laws 
and Newtonian concepts should be introduced as derivatives from these more 
profound theories, as they really are, and the limitations of the Newtonian
concepts must be clearly stressed. 

Let us make two final remarks. Contrary to popular belief, it seems  Einstein 
himself never used $E=mc^2$ in the context of the equivalence of energy and 
mass --- only $E_0=mc^2$, that is, equivalence of the rest energy and the 
invariant mass \cite{3,47,48}. It may seem tempting to use this fact as 
evidence against velocity-dependent relativistic mass. However, in our 
opinion, this fact is completely irrelevant in the context of the present 
article where we appeal not to Einstein's authority but to the logic of 
special relativity.

It may seem also surprising that we do not mention the relation of mass to 
gravity. However, there is a good reason for this. General relativity provides 
another drastic change in our concept of mass, deserving its own story.
It is true that the Newtonian concept of gravitational mass can be 
relativistically generalized in some simple situations. For example, 
if a heavy object with mass $M$ moves at relativistic velocity past to 
a test particle initially at rest with a large enough impact parameter, 
it induces a change in the test particle's transverse velocity corresponding 
to the gravitational mass of the moving body, $\gamma(1+\beta^2)M$, not 
$\gamma M$ \cite{49}. Again, this fact should not be used as an argument 
against relativistic mass $\gamma M$. Instead we should be aware of the 
dramatic changes that general relativity requires of our Newtonian intuition. 
It turns out that it is impossible to give a general definition of a system's 
total mass in general relativity. Even for isolated systems, which produce 
asymptotically flat spacetimes, two reasonable definitions of the total mass 
can be envisaged, related to the Arnowitt-Deser-Misner and Bondi 
energy-momentum tensors at spatial infinity, respectively \cite{50}. We do 
not pursue these subtle matters here any further; again, 
the concept of mass in general relativity deserves its own story.    

\section*{acknowledgments}
The work is supported by the Ministry of Education and
Science of the Russian Federation and in part by Russian Federation President 
Grant for the support of scientific schools NSh-2479.2014.2 and by 
RFBR grant 13-02-00418-a. The author is indebted to Daniel V. Schroeder for 
editing the manuscript.


\begin{thebibliography}{99}
\bibitem{1}
C.~G.~Adler,
``Does mass really depend on velocity, dad?,''
Am.\ J.\ Phys.\  {\bf 55}, 739--743 (1987).

\bibitem{2}
L.~B.~Okun, 
``The Concept of Mass,''
Phys.\ Today {\bf 42} (6), 31--36 (1989).

\bibitem{3}
L.~B.~Okun, 
``Mass versus relativistic and rest masses,''
Am.\ J.\ Phys.\  {\bf 77}, 430--431 (2009).

\bibitem{4}
L.~B.~Okun,
``The concept of mass in the Einstein year,''
arXiv:hep-ph/0602037, 2006. 

\bibitem{5}
J.~Roche,
``What is mass?''
Eur.\ J.\ Phys.\  {\bf 26}, 225--242 (2005).

\bibitem{6}
G.~Oas, 
``On the Abuse and Use of Relativistic Mass,'' 
arXiv:physics/0504110, 2005.

\bibitem{7}
Z.~K.~Silagadze,
``Relativity without tears,''
Acta Phys.\ Polon.\  B {\bf 39}, 811--885 (2008),
arXiv:0708.0929 [physics.ed-ph], 2007.

\bibitem{8}
W.~Rindler's letter in ``Putting to rest mass misconceptions,''
Phys.\ Today {\bf 43} (5), 13 (1990). 

\bibitem{9}
W.~Rindler, {\it Relativity: Special, General and Cosmological}
(Oxford University Press, Oxford, 2006).

\bibitem{10}
T.~R.~Sandin, 
``In defense of relativistic mass,''
Am.\ J.\ Phys.\  {\bf 59}, 1032--1036 (1991).

\bibitem{11}
A.~A.~Ungar,
\textit{Hyperbolic Triangle Centers: The Special Relativistic Approach}
(Springer-Verlag, New York, 2010).

\bibitem{12}
M.~Jammer,
\textit{Concepts of Mass in Contemporary Physics and Philosophy}
(Princeton University Press, Princeton, 2000).

\bibitem{12P}
A.~A.~Ungar,
``When relativistic mass meets hyperbolic geometry,'' 
Commun.\ Math.\ Anal.\ {\bf 10} (1), 30--56 (2011).

\bibitem{13}
R.~Penrose, \textit{The Road to Reality. A Complete Guide to the Laws of the 
Universe} (Alfred A. Knopf, New York, 2005), p. 434.

\bibitem{14}
L.~B.~Okun,
``Formula $E=mc^2$ in the year of physics,''
Acta Phys.\ Polon.\  B {\bf 37}, 1327--1332 (2006).

\bibitem{15}
J.-M.~L\'{e}vy-Leblond,
``On the Nature of Quantons,''
Science \& Education {\bf 12}, 495--502 (2003).

\bibitem{16}
J.-M.~L\'{e}vy-Leblond,
``L'\'{e}nergie apr\'{e}s Einstein (pour comprendre ``eu \'{e}gale emme 
c\'{e}-deux''),''
Bull.\ Union des Physiciens {\bf 769}, 1--14 (1994).

\bibitem{17}
C.~F.~von Weizs\"{a}cker,
``Physics and philosophy,''
in J.~Mehra (ed.), \textit{The Physicist's Conception of Nature} (D.~Reidel 
Publishing Company, Dordrecht, 1973), pp. 736--746.
               
\bibitem{19}
L.~B.~Okun,
``The `Relativistic' Mug,''
arXiv:1010.5400 [physics.pop-ph], 2010.

\bibitem{20}
L.~D.~Landau and E.~M.~Lifshitz, \textit{The Classical Theory of Fields}
(Pergamon Press, Oxford, 1971). 

\bibitem{21}
R.~M.~Marinho, Jr.,
``Noether's theorem in classical mechanics revisited,''
Eur.\ J.\ Phys.\  {\bf 28}, 37--43 (2007), \\
arXiv:physics/0608264, 2006.

\bibitem{22}
J.~Baez,
``Symmetry under boosts gives what conserved quantity?,'' 
http://www.math.ucr.edu/home/baez/boosts.html

\bibitem{23}
P.~Olver, \textit{Applications of Lie Groups to Differential Equations}
(Springer, New York, 1993), p. 279.

\bibitem{24}
D.~Tong,
``Lectures on Quantum Field Theory,''
http://www.damtp.cam.ac.uk/user/tong/qft.html

\bibitem{25}
S.~Coleman, 
``Quantum Field Theory'' (handwritten lecture notes for Coleman's course 
transcribed by B.~Hill), part 1, 
http://www.damtp.cam.ac.uk/user/tong/qft/col1.pdf

\bibitem{26}
J.~A.~De Azc\'{a}rraga and J.~M.~Izquierdo,
\textit{Lie Groups, Lie Algebras, Cohomology and Some Applications in 
Physics} (Cambridge University Press, Cambridge, 1995).

\bibitem{27}
L.~D.~Landau and E.~M.~Lifshitz, \textit{Mechanics}
(Butterworth-Heinemann, Oxford, 2004).

\bibitem{28}
J.-M.~L\'{e}vy-Leblond,
``Group-theoretical foundations of classical mechanics: The Lagrangian 
gauge problem,''
Commun.\ Math.\ Phys.\  {\bf 12}, 64--79 (1969).

\bibitem{29}
G.~Marmo, G.~Morandi, A.~Simoni, and E.~C.~G.~Sudarshan,
``Quasiinvariance and central extensions,''
Phys.\ Rev.\  D {\bf 37}, 2196--2205 (1988).

\bibitem{29P}
S.~Garc\'{i}a,
``Hidden invariance of the free classical particle,''
Am.\ J.\ Phys.\  {\bf 62}, 536--544 (1994), arXiv:hep-th/9306040, 1993.

\bibitem{30}
L.~D.~Faddeev,
``Operator Anomaly For The Gauss Law,''
Phys.\ Lett.\  B {\bf 145}, 81--84 (1984).

\bibitem{30A}
E.~Zeidler,
\textit{Quantum Field Theory III: Gauge Theory} (Springer-Verlag,
Berlin, 2011), pp. 1009-1026.

\bibitem{30B}
P.~Bamberg and S.~Sternberg, 
\textit{A Course in Mathematics for Students of Physics, Vol. 2}
(Cambridge University Press, Cambridge, 1999), pp. 407-457.

\bibitem{30C}
D.~C.~Isaksen,
``A Cohomological Viewpoint on Elementary School Arithmetic,''
Am.\ Math.\ Mon.\  {\bf 109}, 796-805  (2002).

\bibitem{31}
V.~Bargmann,
``On Unitary Ray Representations of Continuous Groups,''
Annals Math.\  {\bf 59}, 1--46 (1954).

\bibitem{32}
E.~P.~Wigner,
``On Unitary Representations of the Inhomogeneous Lo\-rentz Group,''
Annals Math.\  {\bf 40}, 149--204 (1939). Reprinted in
Nucl.\ Phys.\ Proc.\ Suppl.\  {\bf 6}, 9--64 (1989). 

\bibitem{33}
H.~C.~Ohanian,
``Einstein's $E=mc^2$ mistakes,''
arXiv:0805.1400v2 [physics.hist-ph], 2008.

\bibitem{34}
G.~M.~Tuynman and W.~A.~J.~Wiegerinck,
``Central extensions and physics,''
J.\ Geom.\ Phys.\  {\bf 4}, 207--258 (1987).

\bibitem{35}
R.~P.~Feynman and A.~R.~Hibbs, \textit{Quantum Mechanics and Path Integrals}
(McGraw Hill, New York, 1965).

\bibitem{36}
J.~Ogborn and E.~F.~Taylor,
``Quantum physics explains Newton's laws of motion,''
Phys.\ Educ.\  {\bf 40}, 26--34 (2005).

\bibitem{37}
J.-M.~L\'{e}vy-Leblond,
``Classical Apples and Quantum Potatoes,''
Eur.\ J.\ Phys.\  {\bf 2}, 44--47 (1981).

\bibitem{38}
A.~Land\'{e},
``Quantum Fact and Fiction IV,''
Am.\ J.\ Phys.\  {\bf 43}, 701--704 (1975).

\bibitem{39}
J.~M.~Houlrik and G.~Rousseaux,
``\thinspace`Nonrelativistic' kinematics: Particles or waves?,'' \\
arXiv:1005.1762 [physics.gen-ph], 2010.

\bibitem{40}
J.-M.~L\'{e}vy-Leblond,
``Quantum fact and classical fiction: Clarifying Land\'{e}'s pseudo-paradox,''
Am.\ J.\ Phys.\  {\bf 44}, 1130--1132 (1976).

\bibitem{41}
J.-M.~L\'{e}vy-Leblond,
``Neither waves, nor particles, but quantons,''
Nature {\bf 334}, 19--20 (1988).

\bibitem{42}
J.-M.~L\'{e}vy-Leblond,
``Galilei group and nonrelativistic quantum mechanics,''
J.\ Math.\ Phys.\  {\bf 4}, 776--788 (1963).

\bibitem{43}
F.~Wilczek,
``The origin of mass,''
Mod.\ Phys.\ Lett.\  A {\bf 21}, 701--712 (2006).

\bibitem{44}
F.~Wilczek,
``Particle physics: Mass by numbers,''
Nature {\bf 456}, 449--450 (2008).

\bibitem{45}
I.~B.~Khriplovich, \textit{General Relativity} (NITs RKhD, Izhevsk, 2001),
p.~57 (in Russian).

\bibitem{46}
E.~C.~G.~Sudarshan and  N.~Mukunda, \textit{Classical Dynamics: A Modern 
Perspective} (John Wiley \& Sons, New York, 1974).  

\bibitem{47}
E.~Hecht, ``Einstein on mass and energy,''
Am.\ J.\ Phys.\  {\bf 77}, 799--806 (2009).

\bibitem{48}
E.~Hecht, ``How Einstein confirmed $E_0=mc^2$,''
Am.\ J.\ Phys.\  {\bf 79}, 591--600 (2011).

\bibitem{49}
D.~W.~Olson and R.~C.~Guarino, ``Measuring the active gravitational mass of 
a moving object,''
Am.\ J.\ Phys.\  {\bf 53}, 661--663 (1985).

\bibitem{50}
R.~M.~Wald, \textit{General Relativity} (University of Chicago Press, Chicago,
1984), chapter 11. 
 
\end{thebibliography}
\end{document}